\documentclass{PoS}
\usepackage{bm,amssymb}

\newcommand{\beq}{\begin{equation}}
\newcommand{\eeq}{\end{equation}}
\newcommand{\bea}{\begin{eqnarray}}
\newcommand{\eea}{\end{eqnarray}}

\newcommand{\order}[1]{{\cal O}(#1)}  
\newcommand{\pslash}[1]{\rlap{/}\kern-0.8pt #1}
\newcommand{\Dslash}{\rlap{/}\kern-2.0pt D}
\def\today{\number\day\space\ifcase\month\or
January\or February\or March\or April\or May\or June\or
July\or August\or September\or October\or November\or December\fi
\space\number\year}
\newcount\mins \newcount\hours
\def\now{\hours=\time \mins=\time
	\divide\hours by60 \multiply\hours by60 \advance\mins by-\hours
	\divide\hours by60 
	\number\hours:\ifnum\mins<10 0\fi\number\mins~}

\def\stacksymbols #1#2#3#4{\def\theguybelow{#2}
    \def\verticalposition{\lower#3pt}
    \def\spacingwithinsymbol{\baselineskip0pt\lineskip#4pt}
    \mathrel{\mathpalette\intermediary#1}}
\def\intermediary #1#2{\verticalposition\vbox{\spacingwithinsymbol
    \everycr={}\tabskip0pt
    \halign{$\mathsurround0pt#1\hfil##\hfil$\crcr#2\crcr
             \theguybelow\crcr}}}

\newcommand{\half}{{\mbox{$\frac{1}{2}$}}}

\makeatother
\renewcommand\d{\partial}

\newcommand\dlr{\raisebox{0.1em}{$\stackrel{\scriptstyle\leftrightarrow}\partial$}}
\newcommand\+{\dagger}

\newcommand\<{\langle}
\renewcommand\>{\rangle}

\newcommand\kF{k_{\mathrm{F}}}
\newcommand\vF{v_{\mathrm{F}}}

\renewcommand\L{\mathcal{L}}

\newcommand{\VEC}[1]{\bm{\mathbf{#1}}} 

\title{Field theoretic study of a cold Fermi gas in the unitary limit}
\ShortTitle{Field theoretic study of a cold Fermi gas in the unitary limit}

\author{Matthew Wingate\\
Institute for Nuclear Theory, University of Washington, 
Seattle, WA 98195-1550, USA
\thanks{address after 1 Sep 2006: DAMTP, Wilberforce Road, 
University of Cambridge, Cambridge CB3 0WA,
UK.}\\
E-mail: \email{M.Wingate@damtp.cam.ac.uk}}  

\abstract{
Trapped and cooled gases of alkali atoms can be manipulated to exhibit a
variety of interesting phenomena.  For example, dilute gases of fermionic
atoms, in 2 hyperfine states, can be cooled to temperatures where they become
superfluid.  An external field can be applied to tune the scattering length
$a$.  When $|a|$ exceeds the interparticle spacing, nonperturbative tools are
needed to study the system theoretically.  The unitary limit, $|a|\to\infty$,
is particularly interesting due to its universality and symmetry.  Lattice
field theory and effective field theory can be used to systematically
calculate properties of this system.  Results are presented for the finite
temperature phase transition and for behavior near zero temperature.
}

\FullConference{XXIVth International Symposium on Lattice Field Theory\\
                July 23-28, 2006\\
                Tucson, Arizona, USA}

\begin{document}
\bibliographystyle{apsrev}

\section{Introduction}
\label{sec:intro}

In this talk I present results from studying a dilute gas of 2 species of
nonrelativistic fermions interacting through a short range attraction.  
The most relevant experimental realization of this
system is the trapping and cooling of fermionic isotopes of alkali atoms, a
vibrant activity in many labs around the world.  These atomic gas
experiments, and the theory which describes them, can
also be considered to be models of a dilute neutron gas.

The Hamiltonian is the sum of kinetic energy operators
and a short range potential
\begin{equation}
H~=~ -\frac{1}{2}\left(\sum_{i=1}^{N_1} \nabla_i^2
\;+\; \sum_{j=1}^{N_2} \nabla_j^2\right) \;+\; \sum_{i=1}^{N_1}
\sum_{j=1}^{N_2} v(|\VEC{r}_i - \VEC{r}_j|) \, .
\label{eq:hamil}
\end{equation} 
This work is concerned with equal populations of the 2 species, $N_1 =
N_2 \equiv N$.  We adopt nonrelativistic units $\hbar = m = 1$ since the
fermion mass $m$, not the speed of light, is the relevant kinematic factor
relating temporal and spatial units.  Realistic potentials are the van der
Waals potential $v(r)\sim (r_0/r)^6$ for atomic gases and the Yukawa potential
$v(r)\sim \exp(-r/r_0)/r$ for a neutron gas.  In either case $r_0$ is the
length scale characteristic of the specific potential.  

As we know, low momentum processes do not probe short distances, so under the
right conditions, the many body physics is the same for atoms and neutrons.
The typical momentum in a fermion gas is the Fermi momentum, defined through
the particle density $n=N/V$ to be $k_F \equiv (3\pi^2 n)^{1/3}$.
Quantitatively then, the details of the potential do not affect the physics
when $k_F r_0 \ll 1$, which we call the dilute regime.  The atomic gases are
indeed dilute; interparticle distances are typically 2-3 orders of magnitude
larger than $r_0$.  Although realistic neutron matter is never in the dilute
regime -- without high densities of about 0.1 GeV per $\mathrm{(fm)}^3$,
neutrons $\beta$-decay to protons -- idealized dilute neutron matter is a
valid theoretical limit of realistic neutron matter and a warm-up for nuclear
matter.

Looking at an effective range expansion of the scattering amplitude,
\begin{equation} 
{\cal A}~=~\frac{1}{-1/a + \frac12 k^2 r_0 + \ldots -ik} 
\label{eq:scatteringAmpl}
\end{equation}
(defined through the $S$ matrix as $S\equiv 1 + ik{\cal A}$) we can see that
the only relevant parameter is the S-wave scattering length $a$ in the dilute
regime. (Higher partial waves are negligible at low energies.)  Note we follow
the convention is that $a<0$ corresponds to the 2-body system having no bound
state.  With $-1 \ll k_F a < 0$ the ground state is a
Bardeen-Cooper-Schrieffer superfluid, while at $0 < k_F a \ll 1$ tightly bound
difermion molecules are the ground state and form a Bose-Einstein condensate
at low temperatures.  Mean field theory provides an accurate description in
both of these extremes.  The case of particular interest to us is $k_F|a| \gg
1$, which happens when a bound state is close to threshold.  Here fluctuations
dominate and mean field theory is of no use.  In particular, when $|a|\to
\infty$, the scattering length is no longer a physical length scale.  The only
length scale left in the problem is $k_F^{-1}$ (or the interparticle spacing
$n^{-1/3}$), giving universal physics.  This is called the unitary regime.

The atomic physics experiments are pristine, versatile environments to
study the dependence of many body physics on scattering length.  The 2 species
of fermions in the traps are 2 hyperfine spin states, which react differently
to an external magnetic field.  By tuning this magnetic field to a Feshbach
resonance, the scattering length can be fine-tuned to infinity.

Large scattering lengths arise in nuclear physics as well.  Low energy
nucleon-nucleon scattering is governed by one-pion exchange, so dimensional
analysis would suggest that the scattering length should be the same size as
the pion Compton wavelength, $1.4$ fm.  However, in fact the $nn$ scattering
length is $-18.5$ fm, and in $np$-scattering $a_s = -23.76$ fm and $a_t =
5.42$ fm.  Thus nature exhibits an order-of-magnitude of fine tuning.  Direct
Monte Carlo calculation from lattice QCD of $NN$ scattering and binding
as a function of quark mass is underway and should pin down how this fine
tuning arises from QCD.  The study of the unitary Fermi gas is relevant for
nuclear physics in the dilute, $|a|\to\infty$, and $m_W\to\infty$ limits.

Recent experiments have cooled atomic gases in the unitary regime to
temperatures near the superfluid transition
\cite{Bartenstein:2004bb,Regal:2004aaa,Zwierlein:2004bb,Kinast:2004bb,Bourdel:2004bb}.
For example, striking evidence of quantized vortices signal superfluidity in a
rotating gas of lithium-6 \cite{Zwierlein:2005bb}.  Here we discuss the
application of techniques familiar to us at this conference toward this
interesting system.

\section{Lattice field theory at $\bm{T_c}$}
\label{sec:lattice}

Since the details of the realistic potential in (\ref{eq:hamil}) are
irrelevant for dilute gases, we can replace it by a local 4-fermion
interaction with a tunable bare coefficient $C_0$.  Then the discretized
Lagrangian is the same as in the attractive 3-dimensional Hubbard model
\begin{equation}
\L ~=~ \psi^\+\left( \d_t - \frac{1}{2\xi}\nabla^2 - \mu\right)\psi
+ \frac{C_0}{2}\left(\psi^\+\psi\right)^2 \,.
\label{eq:latticeL}
\end{equation} 
An attractive interaction corresponds to $C_0 <0$. We have introduced an
anisotropy relating spatial and temporal lattice spacings: $\xi \equiv
b^2/b_t$.  An external source for pairing can be included in the Lagrangian:
$J\psi\psi + \mathrm{h.c.}$
Doing so allows one to compute the condensate $\Sigma \equiv\<\psi\psi\>$ in
finite volume.  The result in the limit of $V\to\infty, J\to 0$ is an order
parameter for the spontaneous breaking of fermion number.  Although the
infinite volume limit must be taken before $J\to 0$ limit, in this work we 
assume $J$ is not small enough to see finite volume effects.

In order to integrate out the fermion degrees-of-freedom, an auxiliary
scalar field $\phi$ is introduced to complete the square
\begin{equation}
\frac{C_0}{2}\left(\psi^\+\psi\right)^2 \to -\frac{1}{2C_0}\phi^2 
-\phi\psi^\+\psi \,.
\end{equation}
The result is a partition function which can be expressed 
as a path integral with a nonnegative integrand:
\begin{equation}
Z ~=~\int\! {\cal D}\phi\; \mathrm{det}\left[ |J|^2 + \tilde{K}^\+
\tilde{K}\right]\, e^{-S_\phi} \,.
\end{equation}
It has always been known that the attractive Hubbard model has no sign
problem, but interest lay in the repulsive version as a model of
superconductivity in electron systems.  For dilute Fermi gases, this
theory provides a complete effective field theory with a 
continuum limit \cite{Chen:2003vy}.  

The 4-fermion coupling $C_0$ can be tuned to obtain any
scattering length $a$ through
\begin{equation}
\frac{1}{C_0\xi} ~=~ \frac{b}{4\pi a} \;-\; \int_\mathrm{BZ}
\frac{d^3 p}{(2\pi)^3}\frac{1}{|\hat{\VEC{p}}|^2 (1 +
|\hat{\VEC{p}}|^2/4\xi)} 
\label{eq:match}
\end{equation}
where $\hat{p}_j = (2/b)\sin(p_j b/2)$.  The matching is
obtained by requiring $2\to 2$ scattering calculated in the 
vacuum from (\ref{eq:latticeL}) to give the scattering amplitude
(\ref{eq:scatteringAmpl}).

\begin{figure}
\begin{center}
\includegraphics[width=0.7\textwidth]{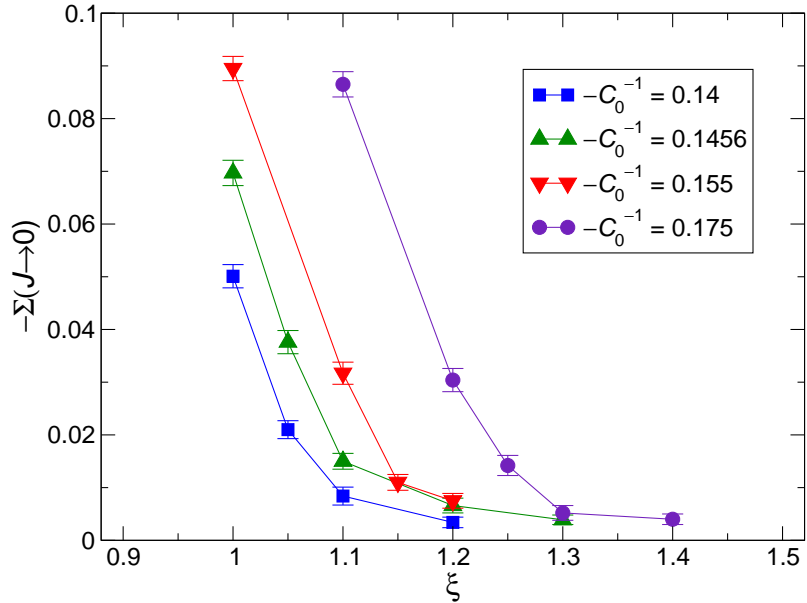}
\end{center}
\caption{The order parameter, after extrapolating $J\to 0$,
as a function of anisotropy for 4 values of the coupling.  A rapid
change in $\Sigma$ is apparent between broken (superfluid) and
restored (normal) phases.}
\label{fig:phasetrans}
\end{figure}

A year ago I performed an exploratory Monte Carlo calculation of the critical
temperature separating the superfluid and normal phases \cite{Wingate:2005xy}.
The lattice volume and chemical potential were fixed to $8^3\times 16$ and
$\mu b^2/\xi = 0.4$ respectively.  As a result the fermion number density was
approximately 1 fermion per 4-5 spatial lattice sites.  Clearly more dilute
calculations are desirable to take the continuum limit. Nevertheless, after
extrapolating the external pairing source $J\to 0$, a clear jump in the
pairing condensate is observed as a function of anisotropy, or temperature.
Figure~\ref{fig:phasetrans} shows the condensate vanishing to zero across the
transition as the U(1) fermion number symmetry is restored.  The four
different curves correspond to four values of the bare coupling.  Since the
matching condition (\ref{eq:match}) also includes the anisotropy $\xi$, the
physical scattering length is varying in the critical region, broadened out
due to finite volume effects.  In order to work at fixed scattering length it
may be more convenient to hold both $\xi$ and $C_0$ fixed and vary $\mu$
across the transition.  A back-of-the-envelope conversion of the critical
region observed in Fig.~\ref{fig:phasetrans} into a critical temperature
yields $T_c/T_F \approx 0.04$ around $1/|a| < 0.5$.

In previous work, the attractive 3D Hubbard model has been studied
in the condensed matter literature \cite{Sewer:2002aaa}, and similar numerical
studies have attempted to model nuclear matter \cite{Muller:1999cp}.

An advance was made recently in Monte Carlo algorithms for this system
\cite{rubtsov:2004rsl,Burovski:2006bb}.  Diagrammatic Monte Carlo
utilizes the convergence of the perturbative expansion for any value of the
coupling, using importance sampling to generate configurations of vertices and
propagators in position space.  The worm updating algorithm in particular
leads to a shortening of autocorrelation times \cite{Prokofiev:2001aaa}.
Calculations have been done using a finite scaling
analysis and have taken the continuum limit.  They find $T_c/T_F =
0.152(7)$~\cite{burovski-2006-96,Burovski:2006bb}.  Note that at finite
lattice spacings they find much lower values for $T_c$, in rough agreement
with my exploratory calculation.  The extrapolation to the continuum limit
relies on correctly fitting the sizable discretization effects.

A calculation using different Monte Carlo methods quotes $T_c/T_F = 0.23(2)$
\cite{Bulgac:2005pj}, while an upper bound $T_c/T_F < 0.14$ was placed in
another work \cite{Lee:2005it}.  The systematic errors appear under tightest
control in the work of Burovski {\it et al.}, but discrepancy between the
various methods needs to be understood and resolved.

\section{Effective field theory at $\bm{T=0}$}
\label{sec:phonon}

(This section presents collaborative work undertaken 
with D.~T.~Son \cite{Son:2005rv}.)

Now let us consider temperatures far below the critical temperature,
to where the only relevant degrees of freedom are the phonons, fluctuations
$\varphi(t,\VEC{x})$ in the phase of the zero temperature condensate
$\<\psi\psi\> =|\<\psi\psi\>|\exp(-2i\varphi)$.  Phonons dominate the physics
when the thermal wavelength is much larger than the coherence length, roughly
the size of the correlated fermion pair: $\sqrt{2\pi/T} \gg v_F/\Delta_0$,
where the Fermi velocity $v_F$ is defined in terms of the fermion number
density and $\Delta_0$ is the gap in the fermion spectrum.  The phonon is a
massless Goldstone boson resulting from the spontaneous breaking of the U(1)
particle number symmetry: $\psi \to \exp(i\alpha)\psi$,
$\psi^\+\to\psi^\+\exp(-i\alpha)$.  Below we construct an effective 
field theory for the phonons in a unitary Fermi gas.

In using symmetries to constrain terms in the phonon Lagrangian,
it is convenient to absorb the chemical potential into a field
$\theta(t,\VEC{x}) \equiv \mu t - \varphi(t,\VEC{x})$.
The lowest-order phonon Lagrangian is 
\begin{equation}
\L_0 ~=~ c_0\left( \d_t\theta - \frac{|\nabla\theta|^2}{2}\right)^{5/2} \,.
\label{eq:phononL-LO}
\end{equation}
Galilean invariance forces the first derivatives of $\theta$ to appear
in the linear combination $X \equiv \d_t\theta - |\nabla\theta|^2/2$.
Scale invariance dictates the functional form of $\L_0$ at the unitary
point $1/a=0$; dimensional analysis permits other powers of $X$ 
when $a$ is finite.  The coefficient $c_0$ must be determined from
experiment or from calculation in the microscopic theory,
for example from the energy per fermion in the ground state.  Below we
write $c_0 = 2^{5/2}/(15 \pi^2 \xi^{3/2})$, where $\xi$ is the ratio of
the energy per particle compared to the ideal Fermi gas energy per particle.

The lowest-order Lagrangian (\ref{eq:phononL-LO}) is equivalent to Landau's
superfluid hydrodynamics.  A definite power counting scheme is needed to go
beyond leading order.  Since $\theta$ never appears undifferentiated, we start
by counting all first derivatives of $\theta$ as terms of $\order{1}$.  Each
further derivative brings in a power of momentum $\d_t^n \sim \d_i^n \sim
{\cal O}(p^n)$.

It is convenient to gauge the U(1), introducing an external field
$(A_0,\VEC{A})$.  We also write the interaction using an auxiliary scalar
field $\sigma$ with 2 free parameters $q_0$ and $r_0$ which can be tuned to
obtain any scattering length, $1/a=0$ in particular.  The microscopic
Lagrangian from which we start is as follows:
\begin{eqnarray}
  \L &=& \frac i2 \psi^\+ \dlr_t\psi
  \;-\; A_0\psi^\+\psi 
  \;-\; \frac{g^{ij}}{2m}(\d_i\psi^\+ \,-\, iA_i\psi^\+)
  (\d_j\psi \,+\, iA_j\psi) \nonumber \\
  && + \; q_0\psi^\+\psi\sigma \; -\; \frac12 g^{ij}\d_i\sigma\d_j\sigma
  \; -\; \frac {\sigma^2}{2r_0^2} \,.
\label{eq:microscopicL}
\end{eqnarray}
Note that $\L$ is written with a nontrivial spatial metric
$g_{ij}$.  These external fields, $A_0, A_i, g_{ij}$, are useful
tools for analyzing the system's symmetry, taking the limit where
$A_i\to 0$ and $g_{ij}\to\delta_{ij}$ to obtain physical results.
We may wish to keep $A_0$ nonzero to include effects of the atomic
trap.  $A_0$ always appears in combination with the chemical potential,
so that $\mu - A_0(\VEC{x})$ is a local effective chemical potential,
e.g.\ 
\begin{equation}
X ~=~ \mu - A_0(\VEC{x}) - \d_t\varphi - \half(D_i\,\varphi)^2 \, .
\end{equation}

The microscopic action $\int \! d t \,d\VEC{x} \sqrt{g}\,\L$ is invariant
under nonrelativistic general coordinate and conformal transformations.
The low energy effective action must be similarly invariant.
The resulting next-to-leading order Lagrangian, after taking $A_i\to 0$
and $g_{ij}\to\delta_{ij}$, is \cite{Son:2005rv}
\begin{equation}
\L_2 ~=~ c_1 \frac{(\nabla X)^2}{\sqrt{X}} \;+\; c_2\sqrt{X}
\left[(\nabla^2\varphi)^2 - 9 \nabla^2 A_0\right] \,.
\label{eq:phononL-NLO}
\end{equation}

Summarizing results obtained with the NLO superfluid Lagrangian,
we find the leading correction to a linear phonon dispersion relation
\begin{equation}
  \omega ~=~ \sqrt{\frac\xi3}\, \vF q\left[1
   \;-\;\pi^2\sqrt{2\xi}\left(c_1+\frac32c_2\right)\frac{q^2}{\kF^2}\right]
\;+\; \order{q^5\ln q}\, . 
\end{equation}
The static density response function is found to be
\begin{equation}\label{chic1c2}
  \chi(q) ~=~ - \frac{\kF}{\pi^2\xi} \left[
    1 \;+\; 2\pi^2\sqrt{2\xi}\left(c_1 - \frac92 c_2\right) 
    \frac{q^2}{\kF^2}\right] \;+\; \order{q^4\ln q}\, .
\end{equation}
And the static transverse response function is
\begin{equation}
  \chi^T(q) ~=~ -9\,c_2 \sqrt{\frac\xi2}\,\kF\,q^2 \;+\; \order{q^4\ln q}\, .
\end{equation}
A dispersion relation requires that $\chi^T(q) < 0$; consequently $c_2$ must
be positive.  The predictivity of the effective field theory is apparent since
the 3 observables over-constrain the 2 free parameters.  Present experiments
are not able to go to low enough temperatures to test these predictions.
However, Monte Carlo calculations should be able to bridge the gap between low
temperatures, where phonons govern the physics, to temperatures around $T_c$,
where experiments can be performed.  Perhaps these calculations will be able
to test the predictions of this effective theory.

The alert reader might have noticed that the logarithmic corrections
due to loop diagrams are suppressed compared to the NLO terms
by another power of $p^2$, in contradistinction to the chiral
logarithms.  In chiral perturbation theory, the logarithms appearing
at NLO arise from interaction terms where the pion field appears
undifferentiated, e.g.\ $(\bm{\pi}\cdot \d_j \bm{\pi})
(\bm{\pi}\cdot \d^j \bm{\pi})$.  Such
terms are not allowed in the (abelian) U(1) effective theory.

Let us close by emphasizing a few features of this work.  It is remarkable
that imposing nonrelativistic general coordinate invariance imposes additional
constraints on the phonon Lagrangian beyond those dictated by Galilean
invariance.  For example, Galilean invariance would relax a linear combination
of operators dictated by general coordinate invariance, so that $\chi^T(q)$
would depend on 3, not 2, low energy constants.  This does not happen with
relativistic theories: any Lorentz invariant term may be made general
coordinate invariant quite simply.  However, the nonrelativistic
transformation is just the $c\to \infty$ limit of the relativistic
transformation. 

One might question the physical justification of requiring the fermion
Lagrangian (\ref{eq:microscopicL}) to have this coordinate invariance -- after
all, the microscopic degrees-of-freedom are atoms whose phenomenologically
successful potentials are not general coordinate invariant.  However, we take
it as a valid first principle that, since the underlying dynamics in atoms are
described by QED and QCD, which are general coordinate invariant theories, the
nonrelativistic remnant of this invariance persists and has predictive,
refutable consequences as discussed above.

Similarly, it is a logical possibility that the conformal invariance we impose
need not be realized by nature. (Then $(\nabla^2\varphi)^2$ and $\nabla^2 A_0$
would be linearly independent in (\ref{eq:phononL-NLO}).)  However, the fact
that the free Schr\"odinger equation is scale and conformally invariant
\cite{Hagen:1972pd} hints that the unitary Fermi gas should also possess the
same symmetry.

\section*{Acknowledgments}

This work was 
supported by DOE grant number DE-FG02-00ER41132.

\bibliography{mbw}

\end{document}